\documentclass[aps,preprint,nofootinbib]{revtex4}

\usepackage{graphicx,epsfig}
\usepackage{amsmath,amssymb}
\usepackage{url}
\usepackage{color}

\newcommand{\lsim}{\mathrel{\mathop{\kern 0pt \rlap
  {\raise.2ex\hbox{$<$}}}
  \lower.9ex\hbox{\kern-.190em $\sim$}}}
\newcommand{\gsim}{\mathrel{\mathop{\kern 0pt \rlap
  {\raise.2ex\hbox{$>$}}}
  \lower.9ex\hbox{\kern-.190em $\sim$}}}
\newcommand{\neu}{{\widetilde{\chi}^0}}
\newcommand{\neuo}{{{\widetilde{\chi}^0_1}}}

\newcommand{\chamo}{{\widetilde{\chi}^-_1}}
\newcommand{\chao}{{\widetilde{\chi}^\pm_1}}
\newcommand{\gev}{{\;{\rm GeV}}}

\newcommand{\es}{{\varepsilon}}
\newcommand{\tb}{{\tan\beta}}

\graphicspath{ {Plots/} }

\makeatletter       

\begin{document}

\title{ 
       Impact on the Light Higgsino-LSP Scenario from Physics \\
       beyond the Minimal Supersymmetric Standard Model \\[1mm]
}

\author{Kingman Cheung$^{1,2,3}$,  Seong Youl Choi$^{4}$,
        Jeonghyeon Song$^{3,5}$\\[-4mm]
        \mbox{ }
}

\affiliation{$^1$ Dept. of Physics, National Tsing Hua University,
                  Hsinchu, Taiwan \\
             $^2$ Physics Division, National Center for Theoretical
                  Sciences, Hsinchu, Taiwan\\
             $^3$ Division of Quantum Phases \& Devices,
                  School of Physics,
                  Konkuk University,  Seoul 143-701, Korea\\
             $^4$ Dept. of Physics and RIPC, Chonkuk National University,
                  Jeonju 561-756, Korea\\
             $^5$ Dept. of Physics, University of Wisconsin, Madison,
                  WI 53706, USA
}
\date{\today}

\begin{abstract}
The modest addition of the dimension-5 term $\lambda\,
(\widehat{H}_u\cdot\widehat{H}_d)^2/M$ to the superpotential of the minimal
supersymmetric standard model (MSSM) originated from physics beyond
the MSSM (BMSSM) has a significant impact on the scenario of the
Higgsino-dominated neutralino state being the lightest supersymmetric
particle (LSP). It increases the mass difference between the LSP and
the lighter chargino as well as that between the LSP and the
second-lightest neutralino. This
enhances the LHC discovery potential of the chargino and neutralino
decays, producing more energetic charged leptons or pions than
the decays without the BMSSM corrections. Furthermore, the
coannihilation between the lighter chargino or second-lightest neutralino
and the LSP is reduced substantially such that the LSP mass does
not have to be very heavy. Consequently, an almost pure Higgsino LSP
with its mass $\sim 100\gev$ in the BMSSM can account for all the relic density
of cold dark matter in the Universe unless $\tb$ is too large.
\end{abstract}
\maketitle

{\it Introduction} --
The presence of cold dark matter (CDM) in our Universe is now well established
by the very precise measurement of the cosmic microwave background radiation
in the Wilkinson Microwave Anisotropy Probe (WMAP) experiment \cite{wmap}.
A nominal $3\sigma$ range of the CDM relic density is
\begin{eqnarray}
\label{wm}
 \Omega_{\rm CDM}\, h^2 = 0.105 \;^{+0.021}_{-0.030} \;,
\end{eqnarray}
where $h$ is the Hubble constant in units of $100$ km/Mpc/s.

One of the most appealing and natural CDM particle candidates is provided by
supersymmetric models with $R$-parity conservation \cite{hooper}.
This $R$-parity  conservation ensures the stability of the lightest
supersymmetric particle (LSP) so that the LSP can be CDM.
The LSP is in general the lightest neutralino, a linear combination of
neutral electroweak (EW) gauginos and Higgsinos.
Since the LSP nature depends on its compositions,
its detection can vary a lot.

An interesting scenario is the Higgsino-like LSP, which can arise from
a number of supersymmetry breaking models, \textit{e.g.}, focus-point
supersymmetry models \cite{focus} or whenever the $\mu$ parameter is
much smaller than the Bino and Wino masses \cite{low:mu}. In the minimal
supersymmetric standard model (MSSM) \cite{Haber}, the Higgsino-LSP
scenario implies nearly-degenerate Higgsino states: Coannihilation is
too efficient so that the observed CDM relic density requires a
rather heavy Higgsino state with mass around $1$--$1.2$ TeV\,\cite{HiggsinoCollider}. Moreover,
the mass degeneracies between the LSP  ($\neuo$) and the lighter
chargino/second-lightest neutralino ($\chao/\neu_2$) generate too soft
decay products for detecting the states $\chao$ and $\neu_2$ at the LHC.
Thus, the Higgsino-LSP in the MSSM
posts a difficult scenario at the LHC.

In this Letter we show that the modest addition of a dimension-5
term $\lambda\, (\widehat{H}_u\cdot\widehat{H}_d)^2/M$ to the MSSM
superpotential, a scenario beyond MSSM (BMSSM) \cite{dine},
alleviates the difficulties of the Higgsino-LSP scenario.
It has been discussed that non-renormalizable and high-dimensional operators
in new physics models can yield important consequences
in low energy phenomenology \cite{effectiveSUSY,strumia}.
As shall be demonstrated in the following, this dimension-5 term,
$\lambda\, (\widehat{H}_u\cdot\widehat{H}_d)^2/M$, lifts up the degeneracy between the
states $\neuo$ and
$\chao/\neu_2$ \cite{Nsusy}, thus enhancing the discovery potential of the chargino and
neutralino decays with more energetic charged leptons or pions than
the decays in the MSSM. In addition, the coannihilation
of the LSP with $\chao$ or $\neu_2$ is reduced substantially such that
a light Higgsino-LSP with a mass around $100\gev$
can accommodate the WMAP data on the CDM relic density.

{\it BMSSM} --Albeit many virtues of low energy supersymmetry (SUSY),
fine-tuning in the lightest Higgs boson mass $m_h$ motivates additional
degrees of freedom to the MSSM \cite{Nsusy}.
New interactions beyond the MSSM at the TeV scale $M$ may
be encoded in higher-dimensional operators.
Recently, Dine \textit{et al.} \cite{dine} have shown that the most
general dimension-5 superpotential term for the MSSM Higgs sector
is
\begin{eqnarray}
\label{eq:BMSSM:1}
W_{\rm dim-5}
  &=&
\frac{\lambda}{M}\,\left(\widehat{H}_u\cdot \widehat{H}_d\right)^2\,,
\end{eqnarray}
with the SU(2) contraction $\widehat{H}_u\cdot \widehat{H}_d =
\widehat{H}^+_u\widehat{H}^-_d- \widehat{H}^0_u\widehat{H}^0_d$  for the up-type
and down-type Higgs doublet superfields, $\widehat{H}_u$ and $\widehat{H}_d$,
respectively. This dimension-5
operator has been shown to raise easily the lightest Higgs boson mass
above the LEP bound without loss of naturalness \cite{Casas:2003jx}.

Another dimension-5 operator, which breaks SUSY and affects the Higgs
spectrum, is
\begin{eqnarray}
\label{eq:BMSSM:2}
\int d^2 \theta\,\mathcal{Z} \frac{\lambda}{M}
        (\widehat{H}_u\cdot\widehat{H}_d)^2\;,
\end{eqnarray}
where $\mathcal{Z} = \theta^2 m_{\rm SUSY}$ is the spurion field with the
SUSY breaking scale $m_{\rm SUSY}$ \cite{dine}.
If
$m_{\rm SUSY} \simeq |\mu|$, the correction to $m_h$
comes dominantly from the supersymmetric operator
in Eq.\,(\ref{eq:BMSSM:1}) rather than that
in Eq.\,(\ref{eq:BMSSM:2}). The correction is given to leading order
in the dimensionless parameter $\es \equiv \lambda\, \mu /M$ by
\begin{eqnarray}
\delta m_h^2 &=& \es v^2 \left[ 1+2 s_{2\beta} + \frac{2
(m_A^2+m_Z^2) s_{2 \beta}}{\sqrt{(m_A^2-m_Z^2)^2+4 m_A^2 m_Z^2
s_{2\beta}^2}} \right]
\\ \nonumber
&\simeq& 8 \frac{m_A^2}{m_A^2 - m_Z^2} \, v^2 \,
 \frac{\es}{\tan\beta} + {\cal O}\left(\frac{ \es}{\tan^2\beta}
  \right),
\end{eqnarray}
where $v \approx 246$ GeV is the Higgs vacuum expectation value and the
second expression holds for large $\tan\beta$. For simplicity
we take the CP-conserving framework and decoupling limit of $m_A \gg m_Z$
in the following. Note that the BMSSM correction is
inversely proportional to $\tan\beta$ in contrast to the conventional
radiative corrections. The correction $\delta m_h$ normalized by $m_h$
around the LEP bound is roughly $\delta m_h/m_h
\simeq 20 \,\es/\tan\beta$ for $m_A\gg m_Z$. For $\es = 0.05$ the
correction can be as large as $50\% \;(10\%)$
for $\tan\beta = 2\;(10)$.  Therefore, the LEP bound is easily
satisfied by a \emph{positive} $\es\sim 0.1$.

The interaction terms that involve only the Higgsino fields
($\widetilde{H}_{u,d}$) and Higgs fields ($H_{u,d}$)
are given by
\begin{eqnarray}
{\cal L}_H &=& - \mu \left( \widetilde{H}_u\cdot \widetilde{H}_d\right)
               - \frac{\lambda}{M}
          \biggr[ 2\left( H_u\cdot H_d \right )
                   \left(\widetilde{H}_u\cdot \widetilde{H}_d\right )
          \nonumber\\
          && + 2 \left(H_u\cdot\widetilde{H}_d\right )
          \left( \widetilde{H}_u\cdot {H}_d\right )
+ \left ( \widetilde{H}_u\cdot {H}_d\right )^2
+ \left ( {H}_u\cdot \widetilde{H}_d\right )^2
   \biggr ] + {\rm H.c.}\,.
\end{eqnarray}
%
After EW symmetry and SUSY breaking, the modified neutralino
mass matrix ${\cal M}_N$ in the $\{\widetilde{B},\widetilde{W}^3,
\widetilde{H}_d^0, \widetilde{H}_u^0\}$ basis reads:
\begin{eqnarray}
\label{eq:neutralino_mass_matrix}
 {\cal M}_N = \left ( \begin{array}{cccc}
   M_1 &  0  &  -m_Z s_W c_\beta   &  m_Z s_W s_\beta \\
   0   & M_2 &   m_Z c_W c_\beta   & -m_Z c_W s_\beta \\
-m_Z s_W c_\beta &  m_Z c_W c_\beta &  \frac{\lambda}{M} v^2 s^2_\beta
       & -\mu + \frac{2\lambda}{M} v^2 c_\beta s_\beta \\
 m_Z s_W s_\beta & -m_Z c_W s_\beta &
         -\mu + \frac{2\lambda}{M} v^2 c_\beta s_\beta
       & \frac{\lambda}{M} v^2 c^2_\beta
         \end{array}
 \right ),
\end{eqnarray}
and the modified chargino mass matrix ${\cal M}_C$ in the
$\{\widetilde{W}^-, \widetilde{H}^-\}$ basis reads:
\begin{eqnarray}
 {\cal M}_C = \left( \begin{array}{cc}
   M_2 & { }\hskip 0.2cm \sqrt{2} m_W s_\beta \\
   \sqrt{2} m_W c_\beta & { }\hskip 0.2cm \mu
  - \frac{\lambda}{M} v^2 c_\beta s_\beta
   \end{array}
  \right ),
\end{eqnarray}
where $s_\beta =\sin\beta$, $s_W = \sin\theta_W$, {\it etc.}
To leading order in $\varepsilon$,
the BMSSM effects on the masses of $\neuo$, $\neu_2$,
and $\chao$ are, in the light Higgsino
case ($M_{1,2}\gg m_Z, \mu$), \cite{drees}
\begin{eqnarray}
\label{eq:mn:mc:approx}
m_{\neu_{1,2}}
 &\simeq&
|\mu| \left[ 1 - \frac{v^2}{2 \mu^2}(2s_{2\beta}\pm 1)\,\es \right]
+ {\rm sign}(\mu) \frac{(M_1 c^2_W + M_2 s^2_W) m^2_Z}{2M_1M_2}
    (1\pm s_{2\beta})
\,,
 \nonumber\\
m_\chao
 &\simeq&
|\mu| \left[ 1 - \frac{v^2}{2 \mu^2}s_{2\beta}\, \es \right]
-{\rm sign}(\mu) \frac{m_W^2 s_{2\beta}}{M_2}\,,
\end{eqnarray}
where in the first equation the upper (lower) sign is for $\neuo$ ($\neu_2$) mass.
With increasing $\varepsilon$ both the $\neuo$ and $\chao$ masses
decrease; the $\widetilde{\chi}^0_2$ mass increases
for large $\tan\beta\agt 4$ but decreases for small $\tb \lsim 4$,
if the $m_W/M_{1,2}$ correction is ignored.
However, since the LSP mass drops faster than the $\chao$ mass with
increasing $\es$, sizable mass differences between $\chao$ and $\neuo$
as well as between $\widetilde{\chi}^0_2$ and $\neuo$ are developed.
Due to the decreasing mass of $\chao$ with $\es$,
the lower mass bound of $m_\chao \gsim 94\gev$\,\cite{PDG}
can constrain the BMSSM light Higgsino-LSP scenario.
If $\mu$ is negative, the
second term in the expressions of $m_\chao$ in
Eq.\,(\ref{eq:mn:mc:approx}) slows down the lighter chargino mass,
which leads to larger mass difference between $\neuo$ and $\chao$.
Therefore, the negative $\mu$ case accommodates larger parameter space
to explain all the WMAP data by the Higgsino-LSP.
However we note that the
combined analysis for the anomalous magnetic moment of the muon
as well as $b \to s \gamma$ prefers a positive $\mu$\,\cite{mumm}.
In what follows, therefore,
we take the case of positive $\mu$.
\begin{figure}[b]
\centering
\includegraphics[height=9cm]{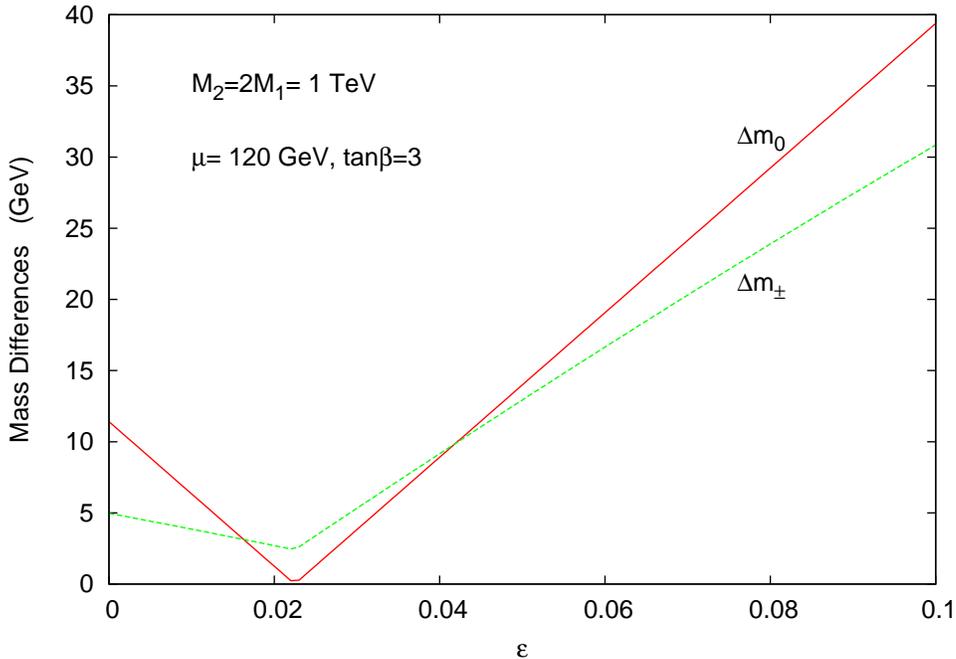}
\caption{\it \small \label{massdiff} The mass difference
         $\Delta m_0=m_{\widetilde{\chi}^0_2}-m_{\widetilde{\chi}^0_1}$
         (solid) between the second lightest neutralino and the LSP and
         the mass difference $\Delta m_\pm = m_{\widetilde{\chi}^\pm_1}
         -m_{\widetilde{\chi}^0_1}$ (dashed)
         between the lighter chargino and the LSP as a function of the
         BMSSM correction parameter $\es$.
}
\end{figure}

{\it Collider detection} --
It is well known \cite{drees} that in the Higgsino-LSP scenario
the mass degeneracy among $\neuo$, $\neu_2$, and $\chao$ renders
their detection at colliders extremely difficult, even though
radiative corrections can increase the mass difference by a few
GeV \cite{radiative:correction}\footnote{The radiative
corrections to the mass splittings can be as large as $10 \gev$
if the stop mixing angle and mass splitting are large. See Ref.\,\cite{Giudice:1995qk}.}.
The decay products of $\chao$ or $\neu_2$
are too soft for their detection.
In the BMSSM, the new contribution
from the dimension-5 term in Eq.\,(\ref{eq:BMSSM:1}) can alleviate the
Higgsino-LSP detection problem significantly by inducing
sizable mass splittings between $\neuo$
and $\chao/\neu_2$.
On the other hand,
the Higgsino fraction of the LSP remains high enough to call the LSP
a pure Higgsino as long as the gaugino masses are large.
The effect of $\es$ on the neutralino mixing matrix $N$,
which diagonalizes the neutralino mass matrix
as $N^*{\cal M}_N N^\dagger ={\rm diag}(m_{\widetilde{\chi}^0_1},
\ldots, m_{\widetilde{\chi}^0_4})$,
corresponds
to the rotation of $\widetilde{H}_d^0$ and $ \widetilde{H}_u^0$ components.
Therefore
the LSP Higgsino fraction, $P_{\widetilde{H}} = |N_{13}|^2 + |N_{14}|^2$,
remains intact by the change of the $\es$ parameter.
In the limit of large gaugino masses,
we have almost pure Higgsino LSP since
$P_{\widetilde{H}} \simeq 1 - \mathcal{O}(m_W/M_1)$.

In Fig.$\,$\ref{massdiff}, we show the mass splittings,
$\Delta m_0\equiv m_{\neu_2}-m_{\neuo}$ and
$\Delta m_\pm\equiv m_{\chao}-m_{\neuo}$, due to the BMSSM
corrections as a function of $\es$ for a
specific choice of parameters: $M_2 = 2 M_1 = 1$ TeV, $\mu=120$ GeV and $\tan\beta=3$.
In this parameter set, the LSP mass is $70-110$ GeV.
For $\es =0.05\,$--$\,0.1$, the mass difference between $\chao $ and $\neuo$
is about $15-30$ GeV and the mass difference between $\neu_2$ and
$\neuo$ is about $15$--$40$ GeV.
These mass splittings
are much larger than those due to radiative corrections, which are
typically a few GeV.
Such sizable mass differences can help us
detect the lighter chargino or second-lightest neutralino by tagging
more energetic charged leptons or jets in the decay products.
Consequently, the phenomenological impact of the BMSSM corrections on
the SUSY search at the LHC is expected significant.
Nevertheless, we do not perform any full-fledged analysis in
  the present work, expecting that such a comprehensive analysis
  will lead to almost the same physical conclusions as those
  described above.

{\it Dark Matter} -- In the MSSM Higgsino-LSP scenario,
the strong coannihilation due to mass
degeneracy pushes the Higgsino mass rather high, about
$1$--$1.2$ TeV, to account for the CDM relic density in Eq.\,(\ref{wm}).
We note, in passing, that the MSSM radiative corrections affect
the relic density rather mildly \cite{nojiri}.
In the BMSSM, however, the large mass differences of $\Delta m_{0}$ and $\Delta m_{\pm}$
can suppress the coannihilation effectively;
a much lighter Higgsino-LSP
can account for the CDM relic density.

Another compelling feature in the BMSSM arises from the modified
$\neuo$-$\neuo$-$Z$ coupling which is proportional to
$(|N_{13}|^2-|N_{14}|^2)$.
The $\es$ dependence can be easily seen from the first row of the
matrix to leading order in the BMSSM corrections, as
\begin{eqnarray}
 \label{eq:N1}
N_{1i} \sim \left(0,~0,~\frac{1+\es_h}{\sqrt{2}},
           ~\frac{1-\es_h}{\sqrt{2}}\right) ,
\end{eqnarray}
where $\es_h = \es v^2 c_{2\beta}/(4 \mu^2)$,
and we ignore small terms of $\mathcal{O}(m_W/M_1)$
as well as an overall phase \cite{diagonalization}.
In the MSSM ($\es=0$), the light Higgsino-LSP scenario implies an almost vanishing
$\neuo$-$\neuo$-$Z$ vertex.
In the BMSSM, the modified neutralino mixing matrix
$N$ in Eq.\,(\ref{eq:N1}) leads to a sizable $\neuo$-$\neuo$-$Z$ vertex,
linearly proportional to $\es$. Therefore, the annihilation
process $\neuo\neuo \to Z \to f \bar{f}$ can be enhanced
by the BMSSM corrections, giving a profound effect on the
relic density of the LSP.

For a simple quantitative estimate we use a useful formula for
the CDM relic density \cite{hooper}
\begin{equation}
\Omega_{\widetilde{H}} h^2
  \approx
\frac{ 0.1\;{\rm pb}}{\langle \sigma_{\rm eff} v \rangle}\;.
\label{usefulformula}
\end{equation}
We include all the $2\to 2$ self-annihilation and coannihilation
processes in calculating the effective annihilation cross section
$\sigma_{\rm eff}$. Since the mass difference $\Delta m_0$ is
substantially larger than $\Delta m_\pm$ for $\es\sim 0.1$ (see
Fig.$\,$\ref{massdiff}), we ignore the $\neuo\neu_2$ coannihilation
in estimating $\langle \sigma_{\rm eff} v \rangle$ and use the following
formula for a crude estimate of the thermally-averaged effective
annihilation cross section, taking into account the coannihilation from
$\widetilde{\chi}^\pm_1$:
 \begin{eqnarray}
\label{eff}
 \langle \sigma_{\rm eff} v \rangle = \frac{
   \sigma_{ \widetilde{\chi}^0_1 \widetilde{\chi}^0_1}\;
   v_{\widetilde{\chi}^0_1\widetilde{\chi}^0_1} +
2\; \sigma_{ \widetilde{\chi}^0_1 \widetilde{\chi}^\pm_1}\;
   v_{\widetilde{\chi}^0_1\widetilde{\chi}^\pm_1} \;
   \left(1 + \frac{\Delta m_\pm}{m_{\widetilde{\chi}^0_1}}\right )^{3/2}\;
   e^{- \Delta m_\pm/T_f}}
  {\left[1 + 2\, \left( 1 + \frac{\Delta m_\pm}{M_{\widetilde{\chi}^0_1}}
   \right)^{3/2}\, e^{-\Delta m_\pm/T_f} \right ]^2 } \;.
\end{eqnarray}
We take the freeze-out temperature $T_f = m_{\widetilde{\chi}^0_1}/25$
and the relative velocity $v_{ij} = 0.3$ during the freeze-out.
For the self-annihilation cross section ($\sigma_{\neuo\neuo}$)
we consider the processes $\neuo\neuo \to hh, Z^0 h,Z^0Z^0, W^+ W^-,
f\bar{f}$. For the coannihilation cross section
($\sigma_{\neuo\chamo}$) we include the processes $\neuo\chamo \to W^-
h^0, W^- Z^0, W^- \gamma, f \bar{f}'$ \cite{Gondolo}. In the following
numerical analysis we fix $m_{\widetilde{Q}_L}=m_{\widetilde{q}_R}=m_{\widetilde{l}}
=-A=m_A =1$ TeV and include all the radiative corrections to the masses of
neutralinos, charginos, and the Higgs bosons.

\begin{figure}[t]
\centering
\includegraphics[scale=1]{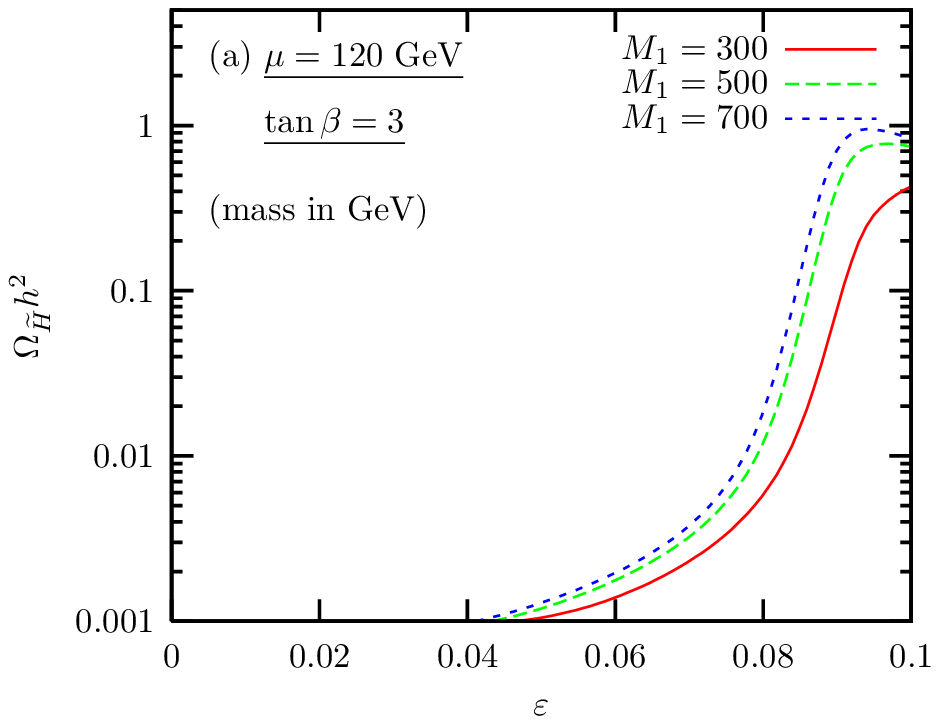}
\\[1cm]
\includegraphics[scale=1]{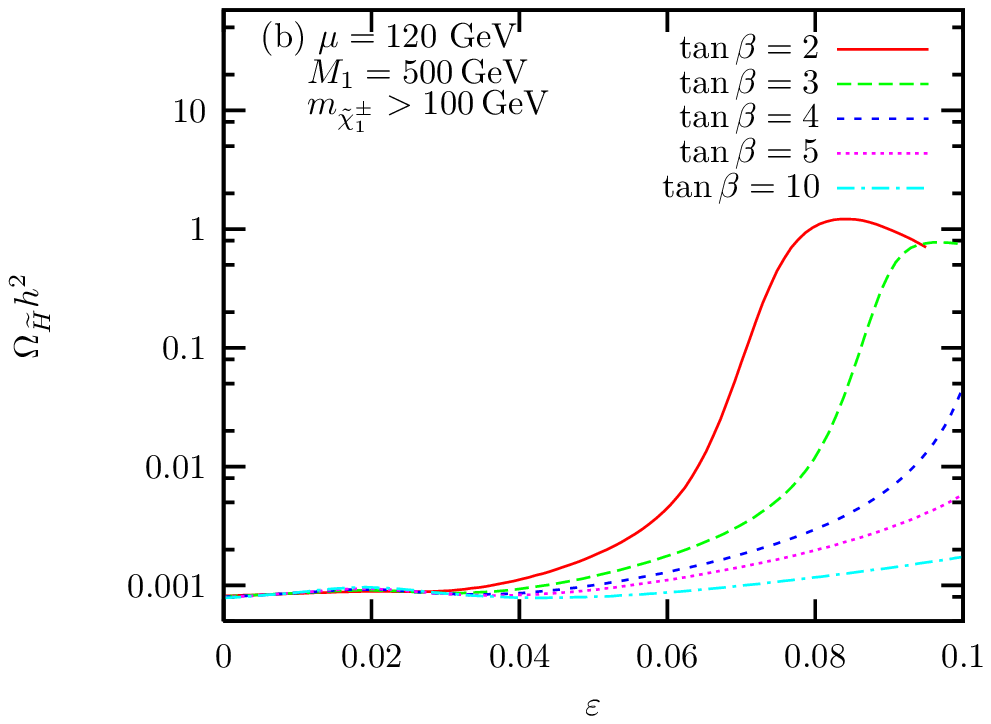}
\caption{\it \small \label{fig:Om:es:varying} The Higgsino-LSP relic
    density $\Omega_{\widetilde{H}} h^2$ as a function of $\es$: (a) for the
    fixed $\tan\beta=3$ with $M_1=300,550,700$ GeV, and (b) for the fixed
    $M_1=M_2/2=300\gev$ with $\tan\beta=2,3,4,5,10$.  We set $\mu=120$ GeV and
    other SUSY mass parameters to 1 TeV.
}
\end{figure}

In Fig.$\,$\ref{fig:Om:es:varying},
we show the $\tan\beta$- and $M_1$-dependence of the relic density
$\Omega_{\widetilde{H}} h^2$ versus $\es$ with the fixed $\mu=120$ GeV.
In Fig.~\ref{fig:Om:es:varying}(a), we fix $\tb=3$ and take
three typical values of $M_{1}=300,500,700\gev$.
As expected, the relic density increases with $\es$
due to the suppressed coannihilation.
In addition, we observe that decreasing 
$M_1$ reduces the LSP relic density.
This is because the effective annihilation cross section
$\sigma_{\rm eff}$, dominated by the process $\neuo\neuo \to f \bar f$
for sizable $\es$, is enhanced by the stronger Bino-Higgsino mixing
for smaller $M_1$.
Nevertheless the $M_1$-dependence of the relic density is rather mild.

On the other hand,
the $\tb$-dependence of $\Omega_{\widetilde{H}} h^2$ is strong,
as can be seen from Fig.~\ref{fig:Om:es:varying}(b).
We take $\tb=2,3,4,5,10$ for the fixed $\mu=120\gev$ and $M_1=500\gev$.
The $\tb=2$ case has the curve terminated at a large $\es$ as
$m_{\widetilde{\chi}^\pm_1}$ gets below the experimental
bound, $m_{\widetilde{\chi}^\pm_1} > 100$ GeV.
Up to $\es \approx  0.08$, the relic density is increasing with $\es$,
because of more suppressed coannihilation.
And  $\Omega_{\widetilde{H}}$ is larger for smaller $\tb$,
which can be attributed to
the $\tb$-dependence of the masses in Eq.\,(\ref{eq:mn:mc:approx}):
Small $\tb\simeq 1$ maximally reduces the LSP mass, and thus enhances the
mass difference $\Delta m_\pm$, suppressing the coannihilation.
The slope of increasing $\Omega_{\widetilde{H}}$ around $\es \simeq 0.08$
is very steep, due to the kinematic closure
of $\neuo\neuo\to W^+ W^-$ mode ($m_\neuo$ is decreasing).
Another interesting feature is that after $\es \approx 0.08$
the relic density in the small $\tb$ case turns its direction and decreases.
This is because, as $\es$ increases,
the $\neuo$-$\neuo$-$Z$ vertex becomes stronger and thus
the self-annihilation
process $\neuo\neuo \to Z \to f\bar{f}$ is enhanced.
Finally we observed that the WMAP data of $\Omega_{\widetilde{H}} h^2\simeq 0.1$
with positive $\mu$ strongly prefer
small $\tb$.
If $\tb \gsim 4$, the WMAP data can be only partially explained.
Of course, if $\mu$ is negative, a much larger parameter space is allowed for the WMAP data.

\begin{figure}[t]
\centering
\includegraphics[height=8cm]{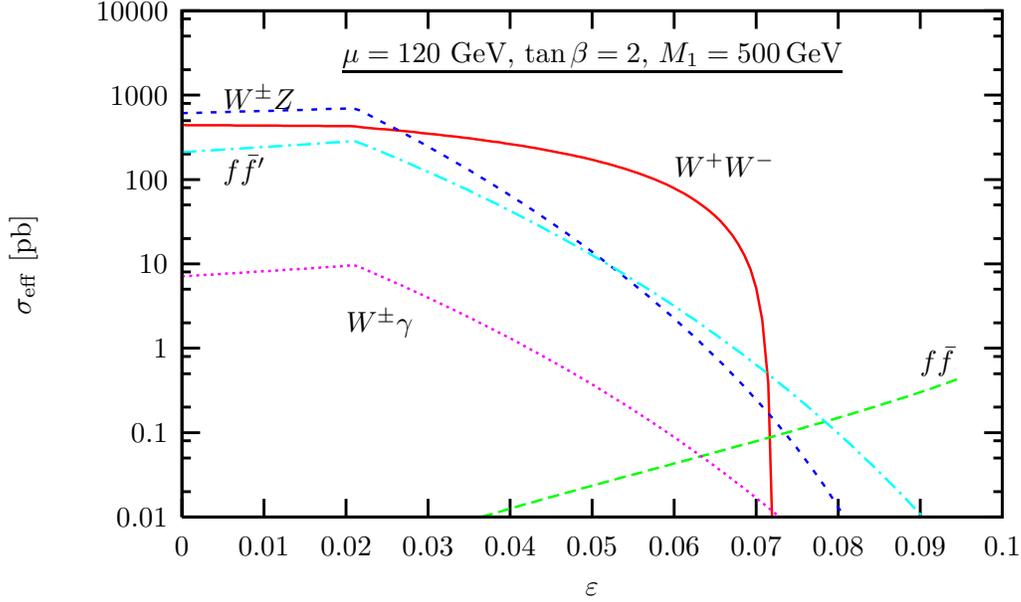}
\caption{\it \small \label{fig:effsig} The effective self and
   coannihilation cross sections. The $W^+ W^-$ mode is from the
   $\neuo\neuo$ self-annihilation while the $W^\pm Z/W^\pm\gamma$ modes
   from the $\neuo\chao$ coannihilation.
  }
\end{figure}

In order to understand the behavior of the relic density against $\es$
(especially for the bumpy shape),
we calculate the contribution of each channel for small $\tb=2$ case.
Figure \ref{fig:effsig} shows the {\it effective} annihilation cross section $\sigma_{\rm eff}$
as a function of $\es$.
Here we set $\mu=120\gev$, and $M_1 = 500\gev$.
We take into account the
thermal suppression factor due to the mass difference. The effective
$\neuo\chao$ coannihilation cross section is defined by
$\sigma_{\rm eff}\,[\neuo\chao] = 2 \sigma_{ \neuo\chao}
\left(1 + {\Delta m_\pm}/{m_{\neuo}}\right)^{3/2}\,
e^{- \Delta m_\pm/T_f}$. For the $\neuo\neuo$ self-annihilation, the
$W^+ W^-$ and $f \bar{f}$ modes are the dominant channels.  As
$\es$ increases, the LSP mass $m_\neuo$ decreases. For $\es\agt 0.072$
the $W^+ W^-$ mode is kinematically closed. The
self-annihilation process $\neuo\neuo\to f \bar{f}$ is enhanced
due to the $\neuo$-$\neuo$-$Z$ vertex being stronger with increasing
$\es$. On the other
hand, the coannihilation channels, dominant for small $\es$,
become suppressed with larger $\es$ due to larger mass splittings.
As a whole, we have the bump-shaped distribution of the relic density
as a function of $\es$ in Fig.\ref{fig:Om:es:varying}(b).
One technical issue can arise when we calculate
the thermally averaged annihilation cross section near $W^+ W^-$
thresholds.
Since we fix the relative velocity,
there is some discrepancy in the
cross section which is sensitive to the actual relative
velocities near the threshold.
A more detailed analysis based on the exact Boltzman
equation will be required for more accurate estimates of the cross
sections, which is, however, beyond the scope of the present short
report.

\begin{figure}[t]
\centering
\includegraphics[width=12cm]{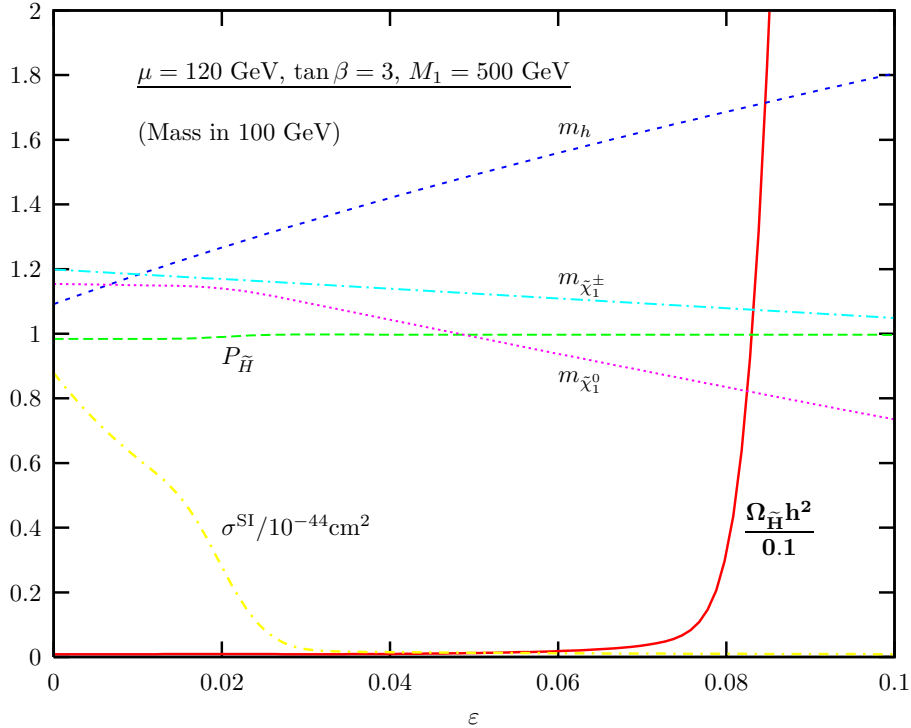}
\caption{\it \small \label{fig:total} As functions of $\es$, we present the
   relic density $\Omega_{\widetilde{H}}\, h^2$ in units of 0.1, the lightest
   CP-even neutral Higgs boson mass $m_h$, the LSP mass,
   and the lighter chargino mass in units of 100 GeV, and the
   Higgsino fraction $P_{\widetilde{H}}$ of the LSP. We fix $\mu=120$ GeV,
   $\tan\beta=3$, and $M_1=500$ GeV. Radiative corrections to the
   neutralino and chargino masses as well as to $m_h$
   are also included.
}
\end{figure}

Another important experimental test for the Higgsino-LSP scenario is
the spin-independent scattering cross section $\sigma^{\rm SI}_{\chi p}$
of the LSP with nucleons. We have calculated the scattering cross section
$\sigma^{\rm SI}_{\chi p}$ based on the input parameters given in
Ref.$\,$\cite{darksusy}. As shown in Fig.$\,$\ref{fig:total}, if $\es$ is
larger than 0.02, the elastic scattering cross section
$\sigma^{\rm SI}_{\chi p}$ is reduced far below the current limits,
mainly because the lightest Higgs boson mass $m_h$ increases with the
BMSSM corrections.
Here we briefly comment on the spin-dependent scattering cross sections.
The spin-dependent scattering process, which is experimentally more
difficult to extract because of the lack of coherent enhancement
unlike the spin-independent process, is mainly mediated by the $Z$
boson. It is  enhanced by the stronger $\neuo$-$\neuo$-$Z$ coupling
with increasing $\es$.

Figure~\ref{fig:total} summarizes all of our findings of the light
Higgsino-LSP scenario in the BMSSM.
For the MSSM parameters, we have chosen moderate values which can explain the WMAP data:
$\mu=120$ GeV, $\tan\beta=3$,
and $M_1=500$ GeV.
We show, as functions of the parameter $\es$, the
relic density $\Omega_{\widetilde{H}}\, h^2$ in units of 0.1, the lightest
Higgs boson mass $m_h$, the LSP mass $m_{\widetilde{\chi}^0_1}$, the
$\chao$ mass (all masses in units of 100 GeV), the Higgsino fraction
$P_{\widetilde{H}}$ of the LSP, and the spin-independent scattering cross
section $\sigma^{\rm SI}_{\chi p}$
in units of $10^{-44}\,{\rm cm}^2$.
As $\es$ increases, the Higgs mass
$m_h$ increases but both $m_\neuo$ and $m_\chao$ decrease.
However, the mass difference $\Delta m_\pm$ increases
while the Higgsino fraction $P_{\widetilde{H}}$ of the LSP stays high $\sim$
98\%. We see a dramatic enhancement of the Higgsino-LSP relic density
$\Omega_{\widetilde{H}}\, h^2$ near $\es=0.08$ or larger,
where $\neuo\neuo\to W^+ W^-$ channel is kinematically closed.
For our parameter choice,
 the light Higgsino-LSP of its mass about 82 GeV, when $\es\approx 0.085$, can
explain all the observed CDM relic density in the Universe. For other
choices of parameters we can still accommodate a light Higgsino-LSP as
a primary candidate for the CDM relic density observed by the WMAP
unless $\tb$ is large.

{\it Conclusions} -- The fine-tuning of the lightest Higgs boson
mass $m_h$ in the MSSM motivates additional interactions beyond the
MSSM around the TeV scale. We have checked that, in the effective
Lagrangian approach, the least-suppressed dimension-5 operator
$\lambda\, (\widehat{H}_u\cdot\widehat{H}_d)^2/M$ added to the MSSM
superpotential can usually increase $m_h$ sufficiently for a moderate
$\tan\beta$ and it can significantly affect the light Higgsino-LSP
scenario. It lifts up the mass degeneracy between the
LSP and the lighter chargino and that between the LSP and the
second-lightest neutralino as much as a few tens of GeV.
As a
result, it is expected to improve significantly the chance of detecting the decay
products of charginos and neutralinos at the LHC. Another important
impact is on the Higgsino-LSP dark matter. Since the mass splittings
suppress the coannihilation processes, the WMAP narrow band for the
CDM relic density can accommodate a light Higgsino-LSP particle of
its mass around $ 100$ GeV for rather small $\tan\beta$.

\acknowledgments{
\noindent
We
  would like to thank Manuel Drees for valuable comments.
The work of KC was supported by the NSC of Taiwan (96-2628-M-007-002-MY3),
the Boost Project of NTHU, and the WCU program through the KOSEF funded
by the MEST (R31-2008-000-10057-0). The work of SYC was supported by
the KRF Grant funded by the Korean Government (KRF-2008-314-C00064).
This work of JS was supported by the Konkuk University.

}

\end{document}